\documentstyle[14lomcon,cite,amsmath,amsfonts,epsfig,graphics]{article}

\begin{document}

\title{SEMILEPTONIC $B$-MESON DECAYS AT $BABAR$}

\author{ Michael Sigamani \footnote{e-mail: sigamani@slac.stanford.edu}}
\address{ Department of Physics, Queen Mary, University of London, \\ Mile End Road, London, E1 4NS, UK}


\maketitle\abstracts{Presented are selected results of semileptonic $B$ decays at $BaBar$.
Two measurements of the Cabibbo-Kobayashi-Maskawa matrix element $|V_{cb}|$ are reported.
One using moments of the hadronic-mass spectrum in inclusive $\bar{B} \to X_{c} \ell^{-} \bar{\nu}$ decays,
and the other, exclusive $\overline{B} \to D \ell^- \bar{\nu}_{\ell}$ decays.
These results are based on data samples of 232 (inclusive $\overline{B} \to X_{c} \ell^{-} \bar{\nu}$)
and 460 (exclusive $\overline{B} \to D \ell^- \bar{\nu}_{\ell}$) million
$\Upsilon(4S) \to B\overline{B}$ decays, recorded by the $BaBar$ detector at the PEP-II $e^+e^-$-storage
rings. Semileptonic $B$ decays are identified by requiring a lepton ($e$ or $\mu$) in events tagged
with a full reconstruction of one of the $B$ mesons in the $B\overline{B}$ pair.}

\section{Introduction}

In the Standard Model of electroweak interactions,
the coupling strength of the $b$ to the $c$ quark in
the weak interaction is described by the Cabibbo-Kobayashi-Maskawa
(CKM)~\cite{Kobayashi:1973fv} matrix element $|V_{cb}|$.
A precise determination of $|V_{cb}|$ is therefore crucial for probing
the CKM mechanism for quark mixing.

Experimentally, $|V_{cb}|$ is obtained using $\bar{B} \to X_{c} \ell^{-} \bar{\nu}$ decays with two distinct approaches:
An exclusive analysis, where the hadronic system $X_{c}$ is reconstructed in a specific mode.
Or, an inclusive analysis where $X_{c}$ is not reconstructed, but rather summed over all possible hadronic final states.
Theoretically, inclusive determinations rely on an Operator Product Expansion
in inverse powers of the $b$-quark mass~\cite{Bigi:1993fe}, which relates the total $\bar{B} \to X_{c} \ell^{-} \bar{\nu}$
rate to $|V_{cb}|$. Whereas exclusive determinations use Form Factors (FF) to describe the hadronization process.
Current measurements of $|V_{cb}|$ using inclusive and exclusive determinations generally
differ by around two standard deviations, with the inclusive result being twice as precise as the exclusive~\cite{pdg}.

\section{Hadronic Reconstruction for the tagged $B$ sample}

The semileptonic sample for both exclusive and inclusive measurements are
selected using $B\overline{B}$ events whereby a full reconstruction
for one of the $B$ mesons is required (referred to as the $B_{tag}$).
The $B_{tag}$~\cite{BrecoVub} candidate is reconstructed using hadronic modes
of the type $\overline{B} \rightarrow D^{(*)} Y$,
where $Y$ represents a collection of charmed hadrons.
The remaining particles in the event are assumed to belong to the
other $B$ (referred to as the $B_{recoil}$), which leaves a
very clean sample of semileptonic events.

\section{Exclusive determination of $|V_{cb}|$ using $\bar{B} \to D\ell^-\bar{\nu}_{\ell}$ decays}

\begin{center}
\begin{figure}[!t]
\includegraphics[scale=0.29]{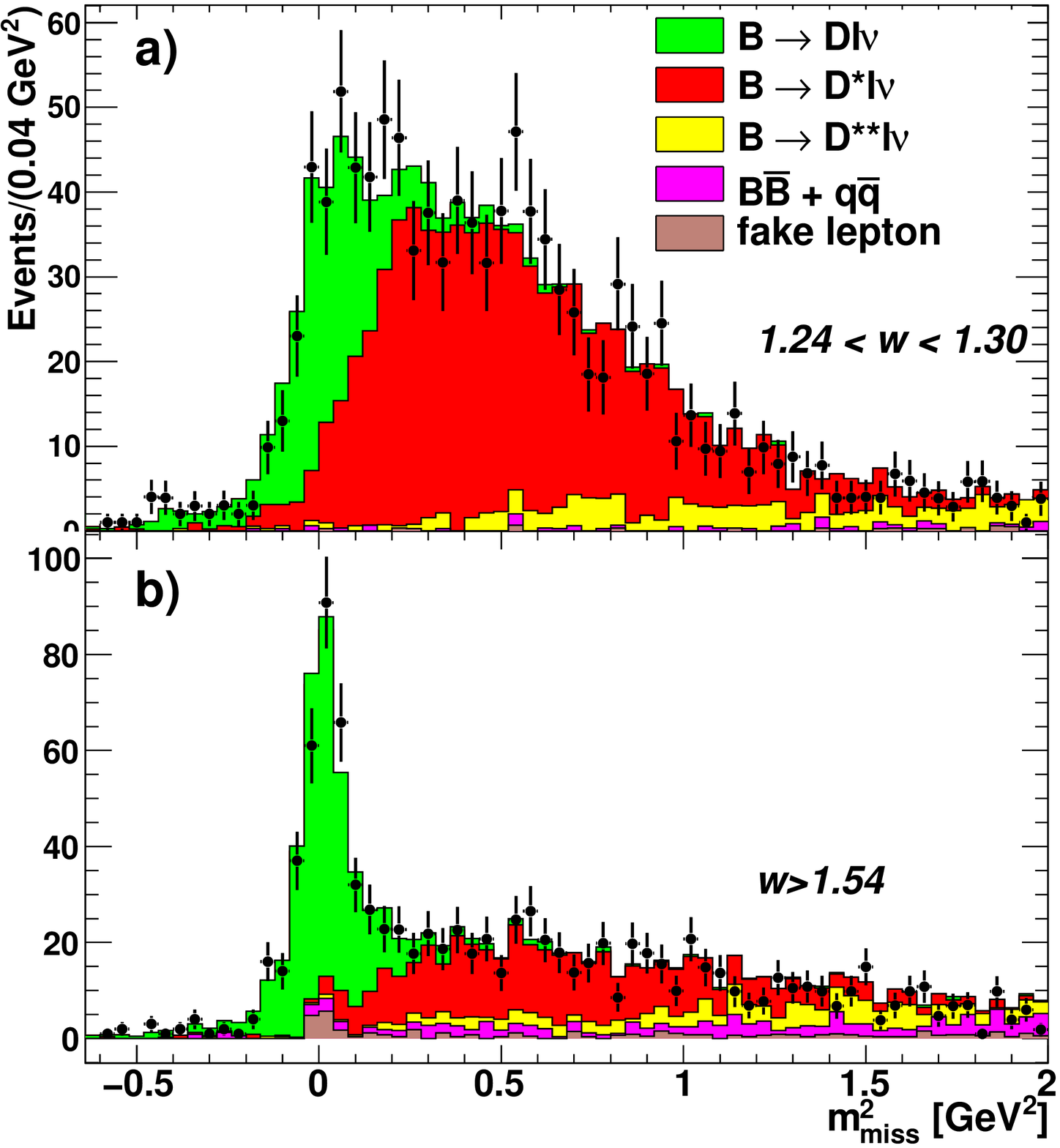}
\includegraphics[scale=0.29]{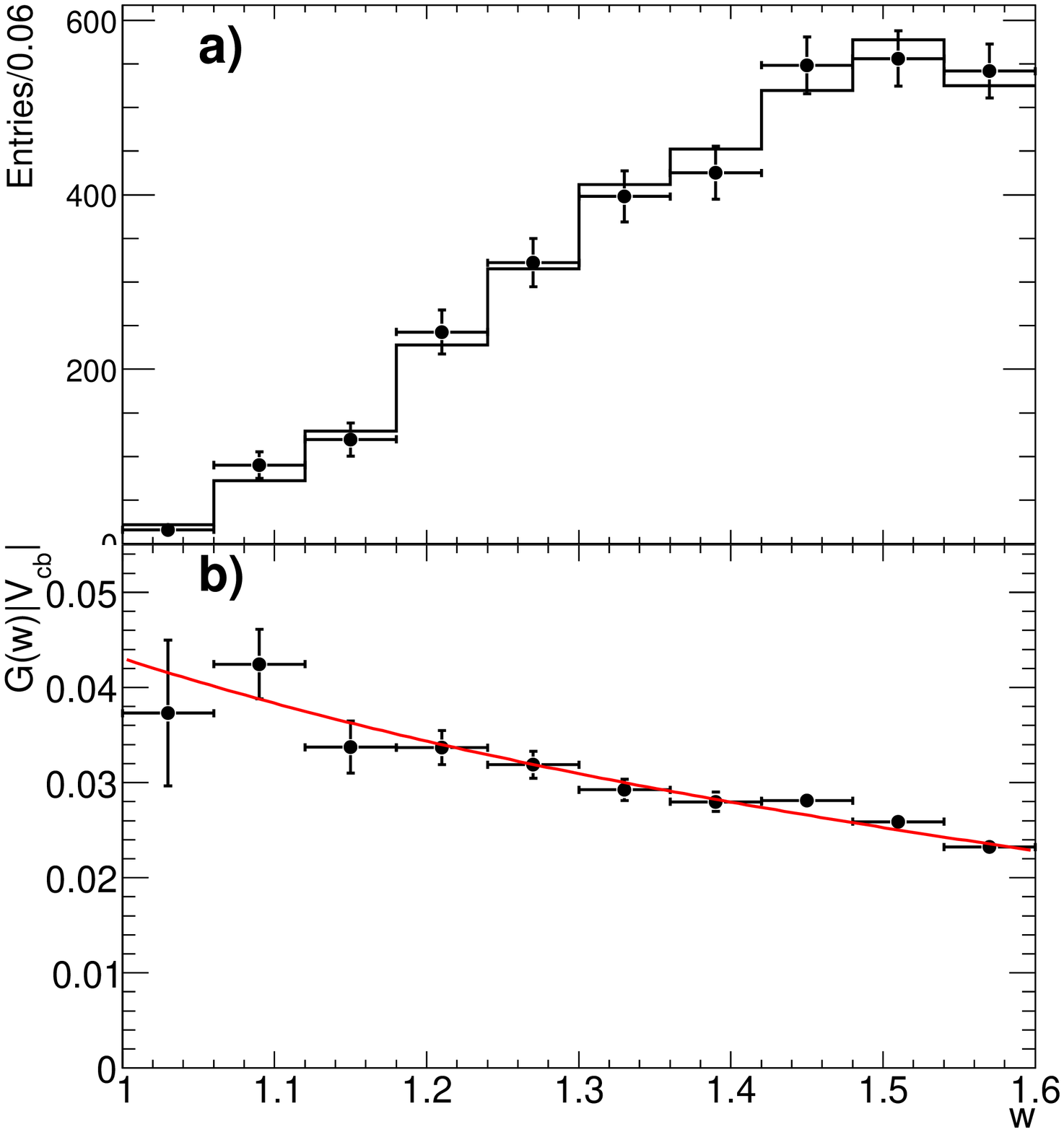}
\caption{Left: Fit to the $m^2_{miss}$ distribution, in two different $w$ intervals,
for $B^- \to D^0 \ell^- \bar{\nu}_{\ell}$: the data (points with error bars) are compared to the results of the overall fit
(sum of the solid histograms). The PDFs for the different fit components are stacked in the order shown in the legend.
Right: (a) The $w$ distribution obtained by summing the  $B^- \to D^0\ell^-\bar{\nu}_{\ell}$
and $\overline{B^0} \to D^+\ell^-\bar{\nu}_{\ell}$ yields. Data (points)
are compared to the results of the overall fit (solid histogram).
(b) The ${\cal G}(w)|V_{cb}|$ distribution corrected for the reconstruction efficiency, with the fit result superimposed.}
\label{fig:mm2Fit}
\end{figure}
\end{center}

The $\bar{B} \rightarrow D\ell^-\bar{\nu}_{\ell}$ decay rate is proportional to the square of $|V_{cb}|$,
and in the limit of very small lepton masses we use the following relation:

\begin{equation}
\label{eq:diffrate_dlnu}                                         
\frac{{\rm d}\Gamma(\bar{B} \to D\ell\nu)}{{\rm d}w}  = \frac{G^2_F}{48 \pi^3 \hbar} M^3_{D} (M_{B}+M_{D})^2 ({w^2-1})^{3/2}  \mid V_{cb} \mid^2 ~ {\cal G}^2 (w),   
\end{equation}

where $G_F$ is the Fermi coupling constant, ${\cal G}(w)$ is a FF which describes the effects of
strong interactions in $\bar{B} \to D$ transitions, and $M_{B}$ and $M_{D}$ are the masses of the $B$
and $D$ mesons respectively. The variable $w$ denotes the product of the $B$ and $D$ meson four-velocities, $V_B$ and $V_D$,
$w = V_B\cdot V_D=(M_{B}^2 + M_{D}^2 - q^2)/(2M_{B} M_{D})$, where $q^2 \equiv (p_{B}-p_{D})^2$, and $p_B$ and $p_D$ are the 
four-momenta of the $B$ and $D$ mesons.
In the limit of infinite quark masses, ${\cal G}(w)$ coincides with the
Isgur-Wise function~\cite{IW}. This function is normalized to unity at zero recoil, where $q^2$ is maximum.
Thus $|V_{cb}|$ can be extracted by extrapolating the differential
decay rate to $w = 1$ using $\overline{B^0} \to D^+ \ell^- \bar {\nu}_{\ell}$ and $B^- \to D^0 \ell^- \bar{\nu}_{\ell}$
decays~\cite{Aubert:2009ac}.
Using the $B_{tag}$ sample, the $B_{recoil}$ is reconstructed by selecting a lepton ($e$ or $\mu$)
with momentum in the CM frame $p_{\ell,{min}}^{*}$ above 0.6 GeV. 
Signal events are then identified using a missing-mass squared value,

\begin{equation} 
m^{2}_{{miss}} = \left[ p(\Upsilon(4S)) -p(B_{{tag}}) - p(D) - p(\ell)\right]^{2}, 
\label{eq:mm2}                                         
\end{equation}

defined in terms of the measured particle four-momenta.
Signal events will peak at zero in the $m^2_{miss}$ distribution as there
is only one associated missing particle (left: Fig.~\ref{fig:mm2Fit}).
Other semileptonic $B$ decays, such as $\bar{B} \to D^{(*,**)} \ell^- \bar{\nu}_{\ell}$ (feed-down)
will yield larger values of $m^2_{miss}$
due to higher numbers of missing particles generated from secondary semi-leptonic decays of the $D^{(*,**)}$.
A measurement of $|V_{cb}|$ is made using a fit to the $w$ distribution,
where data and MC events are evaluated in ten equal-size bins in the interval $1 < w < 1.6$ (right: Fig.~\ref{fig:mm2Fit}).
Data samples are assumed to contain four different contributions:
$\bar{B} \to D\ell^- \bar{\nu}_{\ell}$ signal events,
feed-down from other semileptonic $B$ decays,
combinatorial $B\overline{B}$ and continuum background,
and fake lepton events (predominantly from hadronic $B$ decays with hadrons misidentified as leptons).
From the fit to the combined $\bar{B} \to D \ell^- \bar{\nu}_{\ell}$ sample, a measurement of ${\cal G}(1)|V_{cb}|=(43.0 \pm 1.9 \pm 1.4)\times 10^{-3}$ is made. Using an unquenched lattice calculation, corrected by a factor of 1.007 for QED effects, a value of $|V_{cb}|=(39.8\pm 1.8 \pm 1.3 \pm 0.9_{FF})\times 10^{-3}$ is extracted, where the third error is due to the theoretical uncertainty from ${\cal G}(1)$.

\section{Inclusive determination of $|V_{cb}|$ using hadronic-mass moments in 
$\overline{B} \to X_{c} \ell^{-} \bar{\nu}$ decays}

\begin{figure}[htbp]
\centering
\includegraphics[height=6.2cm]{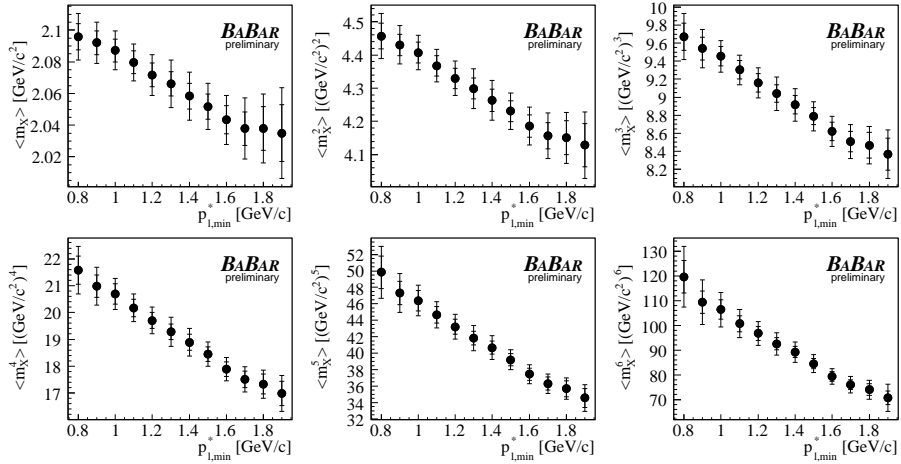}
\caption{Measured hadronic mass moments $\langle m_{X}^{k} \rangle$ with $k = 1...6$ for different
values of $p_{\ell,{min}}^{*}$. The inner error bars correspond to the statistical uncertainties while the full error bars
correspond to the total uncertainties. All the moments are highly correlated since they share subsets of selected events.}
\label{fig:massMoments}
\end{figure}

\vspace{-0.4cm}
We are able to extract $|V_{cb}|$ using measurements of the hadronic mass moments
$ \langle m_{X}^{k} \rangle$~\cite{Aubert:2009qda}, with $k=1,...6$ in semileptonic 
$\overline{B} \rightarrow X_{c} \ell \nu$ decays.
These moments are measured as functions of $p_{\ell,{min}}^{*}$ between $0.8$ GeV/c and $1.9$ GeV/c.
The measured hadronic mass moments $\langle m_{X}^{k} \rangle$ are shown in Fig.~\ref{fig:massMoments}
with $k = 1...6$ as functions $p_{\ell,{min}}^{*}$. The statistical uncertainty consists of
contributions from the data statistics, and the statistics of the MC.
The fit method designed to extract the $|V_{cb}|$ from the moments measurements
has been reported in~\cite{Buchmuller:2005zv}
and is based on a $\chi^2$ minimization technique.
There are eight fit parameters in total namely: $|V_{cb}|$, the quark masses $m_b$ and $m_c$,
the total semileptonic branching fraction ${\cal B}(\overline{B} \rightarrow X_{c} \ell \nu)$, and the dominant
non-perturbative HQE parameters $\mu_{G}^{2}$ and $\rho_{LS}^{3}$.
Combined fits to these moments and moments of the photon-energy spectrum
in $B \rightarrow X_{s} \gamma$ decays~\cite{Aubert:2006qi} have resulted in:
$|V_{cb}| = (42.07 \pm 0.45 \pm 0.70) \times 10^{-3}$.

\vspace{-0.3cm}

\section{Summary and Conclusion}

Presented are measurements of $|V_{cb}|$ using exclusive $\overline{B} \to D \ell^- \bar{\nu}_{\ell}$ decays,
and the moments of the hadronic mass distribution in inclusive $\bar{B} \to X_{c} \ell^{-} \bar{\nu}$ decays.
Using the exclusive determination, the fit to the combined sample
and an unquenched lattice calculation yields: $|V_{cb}|=(39.8 \pm 1.8 \pm 1.3 \pm 0.9_{FF})\times 10^{-3}$.
Using hadronic mass moments in the inclusive determination
yields: $|V_{cb}| = (42.05 \pm 0.45 \pm 0.70) \times 10^{-3}$.

\end{document}